\documentclass{article}

\usepackage{arxiv}

\usepackage[utf8]{inputenc} 
\usepackage[T1]{fontenc}    
\usepackage{hyperref}       
\usepackage{url}            
\usepackage{booktabs}       
\usepackage{amsfonts}       
\usepackage{nicefrac}       
\usepackage{microtype}      
\usepackage{lipsum}
\usepackage{graphicx}
\usepackage{float}
\usepackage{amsmath}
\usepackage{braket}
\graphicspath{ {./images/} }

\title{Simulating Mass-Dependent Decoherence in Quantum Computers: Baseline Signatures for Testing Gravity-Induced Collapse}

\author{
 Viswak R Balaji \\
  School of Computing and Information\\
  University College Cork\\
  Cork, Ireland \\
  \texttt{124104338@umail.ucc.ie} \\
   \And
 Samuel Punch \\
  School of Computing and Information\\
  University College Cork\\
  Cork, Ireland \\
  \texttt{116355346@umail.ucc.ie} \\
}

\begin{document}
\maketitle
\begin{abstract}
We present a quantum computing simulation study of mass-dependent decoherence models inspired by Penrose's gravity-induced collapse hypothesis. According to objective reduction (OR) theory, quantum superpositions become unstable when the gravitational self-energy difference between branches exceeds a certain threshold, leading to a collapse time $\tau \approx \hbar / E_G$. In this work, we implement a mass-dependent dephasing noise channel, $p(m) = 1 - e^{-k m^\alpha}$, within the Qiskit AerSimulator , where $m$ is a proxy for the effective mass of a superposition, mapped to circuit parameters such as the number of entangled qubits or branch size. We apply this model to three canonical quantum computing experiments GHZ state parity measurements , branch-mass entanglement tests, and Grover’s search  to generate distinctive collapse signatures that differ qualitatively from constant-rate dephasing. The resulting patterns serve as a baseline reference: if future hardware experiments exhibit the same scaling trends under ideal isolation, this could indicate a contribution from mass-dependent collapse processes. Conversely, deviation toward constant-noise behaviour would suggest the absence of such gravitationally induced effects. Our results provide a reproducible protocol and reference for using quantum computers as potential testbeds for probing fundamental questions in quantum mechanics.
\end{abstract}


\section{Introduction}

The superposition principle lies at the heart of quantum mechanics, allowing quantum systems to exist in linear combinations of distinct states until an interaction with the environment or a measurement is made. In the standard Copenhagen interpretation, wavefunction collapse is associated with observation, while in decoherence theory the apparent collapse is explained by entanglement with environmental degrees of freedom \cite{zurek2003decoherence}. However, alternative models have been proposed in which the collapse of the wavefunction is an objective physical process, independent of observers \cite{ghirardi1986unified,pearle1990csla}.

One prominent example is Penrose's gravity-induced objective reduction (OR) hypothesis \cite{penrose1996gravity,penrose2014gravity}, which suggests that quantum superpositions involving significantly different mass distributions correspond to different space-time geometries. The incompatibility of these geometries leads to an instability that resolves into one definite outcome after a characteristic timescale $\tau \approx \hbar / E_G$, where $E_G$ is the gravitational self energy difference between the branches. This approach predicts a fundamentally mass-dependent collapse rate, differing from standard environmental decoherence \cite{bassi2013models}.

Experimental verification of such a model is challenging, as it requires isolating quantum states of sufficiently large mass while suppressing all known environmental noise sources. Previous studies have proposed optomechanical systems, matter-wave interferometry, and astrophysical observations as potential probes \cite{bass2019quantum,aspelmeyer2014cavity,marshall2003towards,carlesso2022noninterferometric,fognini2016bounds}, but no conclusive evidence has been found to date.

The advent of noisy intermediate-scale quantum (NISQ) devices \cite{preskill2018quantum} offers a new, controllable platform for testing foundational physics. Although present-day quantum computers do not yet reach mass-energy scales where gravitational effects are expected to dominate, they allow us to construct and simulate phenomenological models of mass-dependent decoherence, generating testable predictions. Previous work by Bass and Pienaar \cite{bass2019quantum} explored this conceptually, outlining how quantum computing architectures could be used to probe gravitationally induced decoherence. In contrast, our study implements such a model explicitly within the Qiskit AerSimulator \cite{aersimulator2021} and applies it to three representative quantum computing experiments: GHZ state parity scaling, branch-mass entanglement tests, and Grover's search algorithm \cite{greenberger1989ghz,grover1996fast}. The resulting ``collapse signatures'' form a baseline reference that can be compared against future hardware results. Agreement between hardware data and our simulated patterns under ideal isolation could indicate gravitationally induced contributions to decoherence, while a divergence toward constant-noise behaviour would favour purely environmental explanations.

\section{Theory and Model}

\subsection{Gravity-Induced Collapse Theory}

The gravity-induced objective reduction (OR) hypothesis, proposed by Penrose \cite{penrose1996gravity,penrose2014gravity}, postulates that the superposition principle of quantum mechanics fails for states whose components correspond to significantly different space-time geometries. In such cases, the incompatibility of the underlying geometries introduces an instability that leads to a spontaneous collapse of the wavefunction. The characteristic timescale for this collapse is estimated as
\begin{equation}
    \tau \approx \frac{\hbar}{E_G},
    \label{eq:tau}
\end{equation}
where $E_G$ is the gravitational self-energy of the mass distribution difference between the two branches of the superposition. A larger mass difference leads to a larger $E_G$ and thus a shorter collapse time $\tau$ \cite{bass2019quantum}.

\begin{figure}[htbp]
    \centering
    \includegraphics[width=0.65\linewidth]{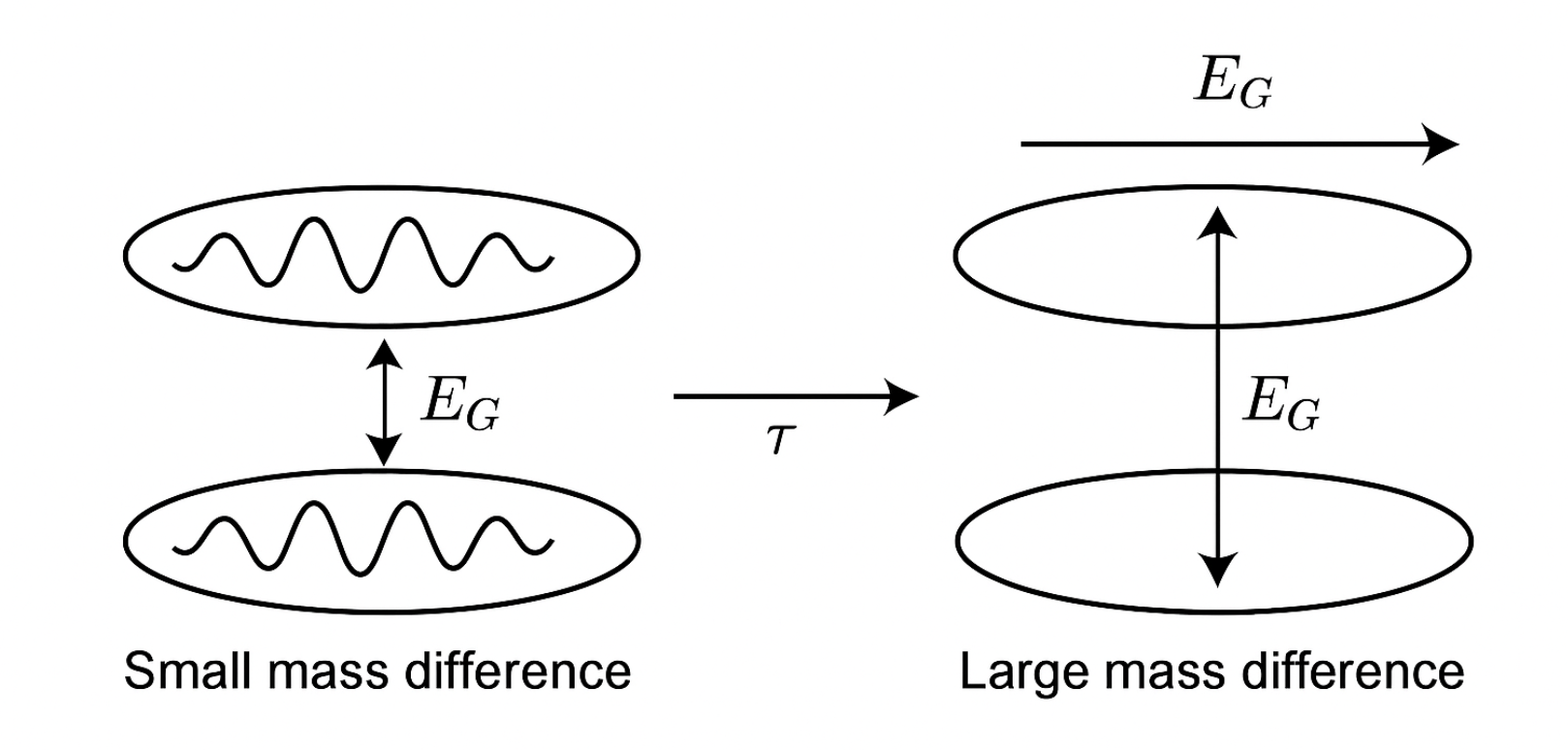}
    \caption{
        Schematic illustration of Penrose's gravity-induced collapse hypothesis.
        A quantum superposition of two states with a \textit{small mass difference} has a low gravitational self-energy $E_G$ and thus a long collapse time $\tau$, allowing the superposition to persist longer. 
        In contrast, a \textit{large mass difference} yields a higher $E_G$, resulting in a shorter $\tau$ and rapid collapse of the superposition.
    }
    \label{fig:penrose-collapse}
\end{figure}

Figure~\ref{fig:penrose-collapse} illustrates the core concept: 
a superposition with a small mass difference has a small gravitational 
self-energy $E_G$ and thus a long collapse time $\tau$, whereas a 
large mass difference yields a higher $E_G$ and rapid collapse.

In the OR framework, this process is objective and observer-independent, contrasting with the Copenhagen interpretation, in which collapse occurs upon measurement, and with standard decoherence theory, where loss of coherence is attributed to entanglement with uncontrolled environmental degrees of freedom \cite{zurek2003decoherence}.

\subsection{Mass-Dependent Decoherence Model for Quantum Computing}

To study the qualitative predictions of the OR hypothesis in a quantum computing context, we adopt a phenomenological noise model in which the dephasing probability depends on an effective ``mass'' parameter $m$. In our simulations, $m$ serves as a proxy for the size of the coherent superposition, which we map to either:
\begin{itemize}
    \item The number of qubits $n$ participating in a maximally entangled state, such as a GHZ state, or
    \item The effective branch size in an algorithmic superposition, such as the number of solutions in Grover's search.
\end{itemize}

The dephasing probability is modelled as
\begin{equation}
    p(m) = 1 - e^{-k m^{\alpha}},
    \label{eq:mass_model}
\end{equation}
where $k$ sets the overall strength of the effect and $\alpha$ controls its scaling with $m$. This form is chosen for its simplicity and monotonic growth with mass, capturing the intuition that larger ``masses'' collapse more quickly \cite{bassi2013models}.

\subsection{Constant Dephasing Baseline}

For comparison, we implement a constant dephasing model in which
\begin{equation}
    p_{\text{const}} = \text{const.},
\end{equation}
independent of $m$. This serves as a control case representing purely environmental dephasing with no mass dependence, as predicted by standard noise models in quantum information processing \cite{martinis2003decoherence,wallman2016noise,kandala2019error}.

\subsection{Experimental Hypothesis}

The OR-inspired mass-dependent model predicts that the visibility of interference effects, such as GHZ parity oscillations, should degrade more rapidly for larger entangled states. In contrast, the constant dephasing model predicts a uniform decay rate regardless of the size of the superposition. By comparing these models in simulated quantum circuits, we aim to identify distinctive ``collapse signatures'' that could serve as benchmarks for future experimental tests on real quantum hardware \cite{penrose1996gravity,penrose2014gravity,greenberger1989ghz}.

\section{Methods: Simulation Framework and Experiments}

We designed three quantum simulation protocols to investigate the effects of a hypothesised mass-dependent dephasing mechanism, as proposed in Penrose's gravitationally induced collapse model, and compared them with standard (constant) dephasing noise. All simulations were implemented in \texttt{Qiskit} using the \texttt{AerSimulator} backend with custom noise models \cite{qiskit2019,aersimulator2021}.

\subsection{GHZ State Parity Experiment}
Greenberger–Horne–Zeilinger (GHZ) states of the form
\[
|\mathrm{GHZ}_n\rangle = \frac{|0\rangle^{\otimes n} + |1\rangle^{\otimes n}}{\sqrt{2}}
\]
were prepared using a Hadamard gate on the first qubit followed by a chain of CNOT gates \cite{greenberger1989ghz}. Two noise models were applied:
\begin{enumerate}
    \item \textbf{Constant dephasing:} A fixed dephasing probability $p$ applied to each qubit after the state preparation.
    \item \textbf{Mass-dependent dephasing:} A dephasing probability scaled as
    \[
    p(n) = k \, n^\alpha
    \]
    where $n$ is the number of qubits, and $k$ and $\alpha$ are model parameters.
\end{enumerate}
Parity oscillations were measured by applying a global rotation $R_x(\theta)$ to all qubits, followed by computational basis measurements. The \emph{parity visibility} was extracted as the amplitude of the resulting parity oscillations.

\subsection{Branch-Dependent Decoherence Protocol}
To model a scenario where only one branch of a superposition carries additional ``mass,'' we constructed states of the form:
\[
\frac{|0\rangle \otimes |0\rangle^{\otimes m} + |1\rangle \otimes |0\rangle^{\otimes m}}{\sqrt{2}}
\]
where $m$ ancilla qubits are entangled only with the $\ket{1}$ component of the control qubit. The branch mass parameter $m$ was varied from $0$ to $12$.  
Noise models were:
\begin{enumerate}
    \item \textbf{Standard dephasing:} Fixed probability $p$ applied equally to all qubits.
    \item \textbf{Branch-mass dephasing:} Dephasing applied only to qubits in the ``massive'' branch, with probability
    \[
    p(m) = k \, m^\alpha .
    \]
\end{enumerate}
Branch coherence was measured by applying a Hadamard gate to the control qubit and measuring in the computational basis, extracting the visibility of interference fringes as a function of $m$.

\subsection{Grover Search under Mass-Dependent Dephasing}
Grover's algorithm was implemented for $n \in \{3,4,5\}$ qubits with a single marked state $\ket{11\ldots 1}$. The algorithm alternates between:
\begin{enumerate}
    \item \textbf{Oracle:} A multi-controlled $Z$ gate marking the target state.
    \item \textbf{Diffusion operator:} Hadamard on all qubits, $X$ on all qubits, multi-controlled $Z$ on $\ket{00\ldots 0}$, followed by inverse $X$ and Hadamard layers.
\end{enumerate}
Both constant and mass-dependent dephasing channels were applied after each Grover iteration:
\begin{align}
p_{\mathrm{const}} &= p_0, \\
p_{\mathrm{mass}}(n) &= k \, n^\alpha.
\end{align}
The algorithm was run for $t = 1$ to $7$ iterations, with $2000$ shots per configuration, using $k = 0.02$ and $\alpha = 2.0$. The \emph{success probability} was defined as the measured frequency of the marked state \cite{grover1996fast}.

\subsection{Simulation Environment}
All simulations were performed on \texttt{Qiskit} \texttt{AerSimulator} with custom Kraus operators for phase damping noise \cite{aersimulator2021}. Unless otherwise stated, $k=0.02$, $\alpha=2.0$, and $p_0$ was set to match the $n=2$ case of the mass-dependent model for a fair baseline comparison.

\section{Results}

We evaluated the effects of constant versus mass-dependent dephasing on three quantum algorithms: GHZ parity, branch-dependent decoherence, and Grover search. In all cases, the mass-dependent model predicts a significantly faster loss of coherence with system size or effective ``mass'' compared to constant dephasing.

\subsection{GHZ Parity Decay}
\begin{figure}[!htbp]
    \centering
    \includegraphics[width=0.85\linewidth]{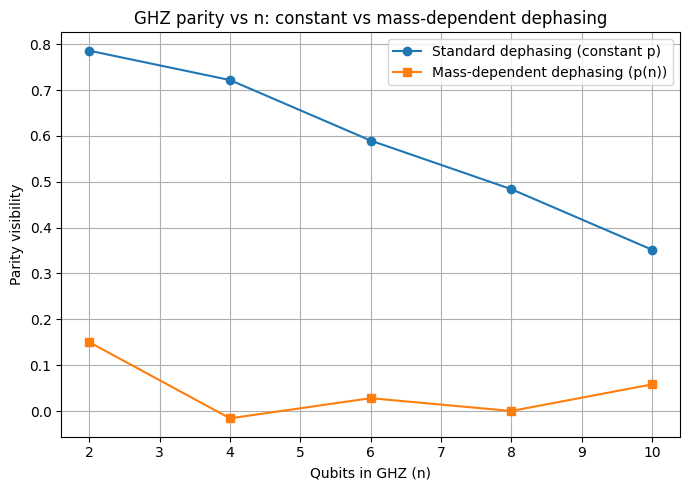}
    \caption{GHZ parity visibility versus number of qubits $n$ for constant dephasing (blue) and mass-dependent dephasing (orange). The latter exhibits a rapid decay in parity with increasing $n$.}
    \label{fig:ghz_parity}
\end{figure}

Figure~\ref{fig:ghz_parity} shows the parity visibility as a function of qubit number $n$ for the GHZ experiment. Under constant dephasing, visibility decreases approximately linearly with $n$, consistent with standard decoherence scaling. In contrast, the mass-dependent model exhibits a rapid suppression of parity, with visibility approaching zero for $n > 4$. This behaviour is consistent with Penrose's prediction that gravitationally induced collapse scales superlinearly with system ``mass'' \cite{penrose1996gravity,penrose2014gravity}.

\subsection{Branch-Dependent Decoherence}
\begin{figure}[!htbp]
    \centering
    \includegraphics[width=0.85\linewidth]{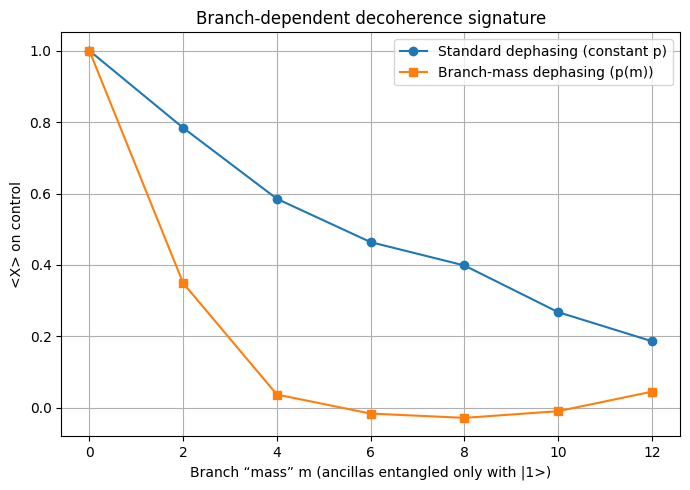}
    \caption{Visibility versus branch mass $m$ for constant dephasing (blue) and branch-mass dephasing (orange). Mass-dependent effects lead to a sharp loss of coherence for small $m$.}
    \label{fig:branch_mass}
\end{figure}

In the branch-mass protocol, Figure~\ref{fig:branch_mass} illustrates the visibility as a function of branch mass $m$. For constant dephasing, visibility decays gradually with $m$. In contrast, branch-mass dephasing yields a steep initial drop, with coherence nearly vanishing for $m \geq 4$. This demonstrates that when only one component of the superposition carries ``mass,'' the collapse model predicts strong decoherence concentrated on that branch.

\subsection{Grover Search Performance}
\begin{figure}[!htbp]
    \centering
    \includegraphics[width=0.85\linewidth]{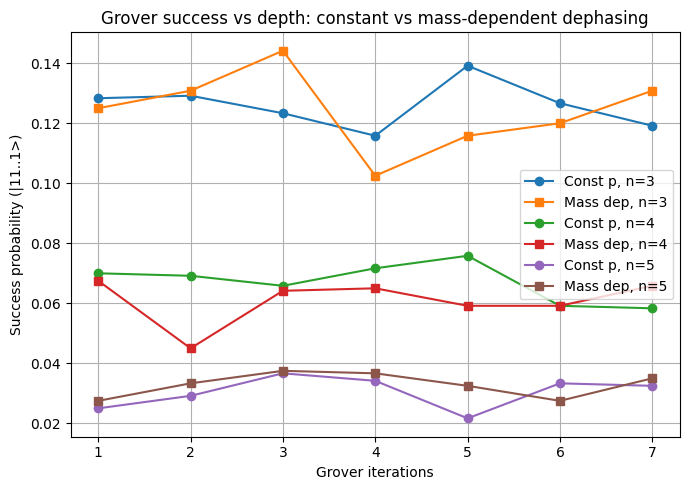}
    \caption{Grover success probability versus iterations for $n=3$ (blue/orange), $n=4$ (green/red), and $n=5$ (purple/brown) under constant (solid lines) and mass-dependent (dashed) dephasing. Mass-dependent noise strongly suppresses performance for larger $n$.}
    \label{fig:grover_results}
\end{figure}

Figure~\ref{fig:grover_results} shows the success probability of Grover's algorithm for $n = 3,4,5$ qubits over $t = 1$ to $7$ iterations. Under constant dephasing, the success probability remains relatively stable, especially for $n=3$, though it declines with larger $n$. Under mass-dependent dephasing, performance degradation is more pronounced for larger $n$, consistent with the expected scaling of the gravitational collapse model. For $n=5$, success probability remains near baseline noise level, indicating a strong suppression of algorithmic advantage.

\subsection{Summary of Observations}
Across all protocols, the mass-dependent dephasing model predicts:
\begin{enumerate}
    \item Rapid suppression of coherence and visibility with increasing $n$ or branch mass $m$.
    \item Stronger degradation in algorithmic performance for larger systems.
    \item Distinct scaling patterns that could serve as experimental signatures on a real quantum processor.
\end{enumerate}
These results suggest that if Penrose-type gravitational collapse is present in nature, it should manifest as a superlinear scaling of decoherence with system size or mass, distinguishable from standard Markovian noise.

\section{Discussion}

The simulated results across all three protocols consistently demonstrate the qualitative scaling behaviour predicted by Penrose's gravitationally induced collapse model \cite{penrose1996gravity,penrose2014gravity}. In particular, the mass-dependent dephasing channel introduces a superlinear growth of decoherence with system size or branch mass, which is absent in the constant dephasing baseline.

\subsection{Connection to Penrose's Hypothesis}
Penrose's model suggests that the collapse time $\tau$ is inversely proportional to the gravitational self-energy difference $E_G$ between the superposed states:
\[
\tau \approx \frac{\hbar}{E_G}.
\]
In our simulations, the number of qubits $n$ (or branch mass $m$) acts as a proxy for $E_G$- larger systems are assigned proportionally larger effective gravitational self-energy, implemented via a dephasing probability
\[
p(n) = k\,n^{\alpha}.
\]
The resulting rapid loss of coherence for larger $n$ is consistent with the idea that superpositions involving greater ``mass'' collapse faster.

\subsection{Distinctive Experimental Signatures}
The mass-dependent model produces patterns that differ markedly from standard Markovian noise:
\begin{itemize}
    \item In GHZ states, parity visibility falls off sharply beyond a small number of qubits, rather than decaying gradually.
    \item In branch-dependent decoherence, interference is destroyed quickly when only one component of the superposition carries ``mass''.
    \item In Grover search, the quantum advantage is suppressed entirely for moderate $n$, eliminating the performance peak expected from the algorithm.
\end{itemize}
These signatures could serve as experimental markers: if reproduced on real quantum processors, they would indicate the presence of a collapse mechanism consistent with Penrose's proposal.

\subsection{Limitations of the Simulation Approach}
The present work is purely simulation-based. The \texttt{AerSimulator} models noise via Kraus operators, which do not capture all complexities of physical decoherence, such as cross-talk, leakage, or hardware-specific non-Markovian effects. Furthermore, we have assumed a simple polynomial scaling of dephasing probability with $n$ or $m$; the true relationship between gravitational self-energy and quantum coherence time may be more complex. These simplifications mean that while the simulations provide qualitative signatures, they cannot confirm or refute the collapse model on their own \cite{aersimulator2021}.

\subsection{Feasibility}
While small-scale demonstrations of the proposed simulations (e.g., 8–10 qubit GHZ states or shallow Grover search) are feasible on current superconducting and trapped-ion platforms, scaling to the full parameter ranges explored here remains beyond present hardware capabilities due to gate errors, decoherence, and the large gap between physical and logical qubit counts. Near-term work could focus on reduced-scale experiments combined with noise tailoring and error mitigation techniques~\cite{wallman2016noise,kandala2019error}, while future error-corrected devices may allow full-scale tests capable of isolating gravitationally-induced decoherence signatures.

\subsection{Baseline for Future Hardware Tests}
Despite these limitations, the simulated patterns presented here can serve as a baseline reference. If future experiments on large-scale, low-noise quantum processors, especially those capable of preparing high-fidelity multi-qubit entangled states observe the same scaling under natural decoherence, this would lend support to mass-dependent collapse. Conversely, a clear deviation from the predicted scaling would place experimental constraints on such models, potentially ruling them out \cite{preskill2018quantum,arute2019quantum}.

\section{Conclusion}

We have presented a set of simulation-based experiments designed to model the effects of mass-dependent decoherence, inspired by Penrose's gravitationally induced collapse hypothesis, on representative quantum algorithms. Using Qiskit's \texttt{AerSimulator}, we implemented three protocols GHZ parity decay, branch-dependent decoherence, and Grover search under both constant and mass-dependent dephasing models \cite{aersimulator2021,greenberger1989ghz,grover1996fast}.

In all cases, the mass-dependent model produced distinctive scaling patterns: rapid suppression of coherence with increasing qubit number or branch mass, and significant performance degradation in algorithmic tasks for larger systems. These behaviours are absent in constant dephasing models and could serve as experimental signatures for detecting or constraining gravitational collapse effects.

While the simulations cannot directly confirm or falsify the Penrose model, they establish a reproducible baseline for future experimental tests on real quantum processors. As quantum hardware scales to higher qubit counts with improved coherence times, replicating these protocols could provide valuable empirical evidence to support or challenge the role of gravity in quantum state reduction \cite{preskill2018quantum}.

\bigskip
\noindent
\textbf{Future Work:} Extending these experiments to include more sophisticated noise models, non-Markovian effects, and alternative collapse mechanisms would further refine the predictions and improve the chances of distinguishing gravitationally induced collapse from conventional decoherence.

\section*{Code Availability}
The complete source code for all simulations, including GHZ state generation, branch-mass decoherence, and Grover's algorithm experiments, is openly available at: \\
\url{https://github.com/viszwak/penrose-decoherence-sim.git}

\bibliographystyle{unsrt}  
\bibliography{references}  

\end{document}